\documentclass[conference]{IEEEtran}
\usepackage{enumerate}
\usepackage{graphicx}
\usepackage{epsf}
\usepackage{subfigure}
\usepackage{amsmath}
\usepackage{amssymb}
\usepackage{array}
\usepackage{setspace}
\usepackage[amsmath,thmmarks,amsthm]{ntheorem}
\usepackage[ruled,lined]{algorithm2e}
\usepackage{algorithmic}
\usepackage{color}
\usepackage{amsfonts}
\usepackage{dsfont}
\usepackage{xfrac}
\usepackage{breqn}
\usepackage{bbm}
\usepackage{cite}
\usepackage{makecell}
\usepackage[english]{babel}
\usepackage{mathtools}
\usepackage{mathrsfs}

\graphicspath{{fig/}}

\theoremseparator{.}

\begin{document}
\title{\Huge Distributed Multi-Agent Deep Q-Learning for Fast Roaming in IEEE 802.11ax Wi-Fi Systems}

\author{\IEEEauthorblockN{Ting-Hui Wang, Li-Hsiang Shen, and Kai-Ten Feng}\\
\IEEEauthorblockA{Institute of Artificial Intelligence innovation, Industry Academia Innovation School \\Department of Electronics and Electrical Engineering  \\
  National Yang Ming Chiao Tung University, Hsinchu, Taiwan\\ 
Email: tina10435641.11@nycu.edu.tw, gp3xu4vu6.cm04g@nctu.edu.tw, ktfeng@nycu.edu.tw}}

\maketitle

\begin{abstract}
    The innovation of Wi-Fi 6, IEEE 802.11ax, was be approved as the next sixth-generation (6G) technology of wireless local area networks (WLANs) by improving the fundamental performance of latency, throughput, and so on. The main technical feature of orthogonal frequency division multiple access (OFDMA) supports multi-users to transmit respective data concurrently via the corresponding access points (APs). However, the conventional IEEE 802.11 protocol for Wi-Fi roaming selects the target AP only depending on received signal strength indication (RSSI) which is obtained by the received Response frame from the APs. In the long term, it may lead to congestion in a single channel under the scenarios of dense users further increasing the association delay and packet drop rate, even reducing the quality of service (QoS) of the overall system. In this paper, we propose a multi-agent deep Q-learning for fast roaming (MADAR) algorithm to effectively minimize the latency during the station roaming for Smart Warehouse in Wi-Fi 6 system. The MADAR algorithm considers not only RSSI but also channel state information (CSI), and through online neural network learning and weighting adjustments to maximize the reward of the action selected from Epsilon-Greedy. Compared to existing benchmark methods, the MADAR algorithm has been demonstrated for improved roaming latency by analyzing the simulation result and realistic dataset.  
\end{abstract}

\section{Introduction}

    The rapid advancements in Wi-Fi technology have been identified as a key enabler for the large-scale deployment of the Internet of Things (IoT), paving the way for an intelligent information society\cite{1}. As Wi-Fi is anticipated to facilitate emerging services and applications, the increased usage of Wi-Fi devices raises concerns about network congestion\cite{1, 2}. Moreover, the COVID-19 pandemic has accelerated the adoption of unmanned intelligent technologies across various sectors, consequently increasing the demand for wireless data services in scenarios such as Smart Warehouses, self-service stores, and automated guided vehicles (AGVs).

As the global population and the number of Wi-Fi devices continue to grow, the demand for low latency and quality of service (QoS) in 6G wireless communication networks has intensified \cite{acm}. Traditional standards and techniques primarily focus on enhancing throughput and optimizing resource allocation using convex optimization or other mathematical analysis methods \cite{3,4}. Recent studies have explored Wi-Fi roaming systems and commercial smart warehouse applications to improve efficiency in commercial settings \cite{5,6}. However, these works do not address channel contention and the time complexity arising from extensive calculations.

Uplink OFDMA-based random access (UORA) contention mechanisms have been analyzed in \cite{7,8,9,m3}, revealing their potential to support multi-user transmissions simultaneously in Wi-Fi 6 scenarios. Despite this, UORA still faces issues with high collision probability, low efficiency, and increased transmission packet delays during the roaming process under high contention. Existing standards like IEEE 802.11k/v/r have aimed to minimize interruptions when Wi-Fi stations (STAs) switch between access points (APs) in the network \cite{10,11}. However, many commercial devices do not support these standards due to hardware limitations, thereby reducing their overall impact. Furthermore, 802.11k/v/r relies on a centralized algorithm design, concentrating control schemes on the AP side and increasing the operational burden on APs. As a result, distributed multi-agent schemes and deep Q-networks (DQNs) have shown potential in dynamically addressing these issues\cite{12, 13}.

    \begin{figure}
    \centering
    \setlength\belowcaptionskip{-1.3\baselineskip}
    \includegraphics[width=0.5\textwidth]{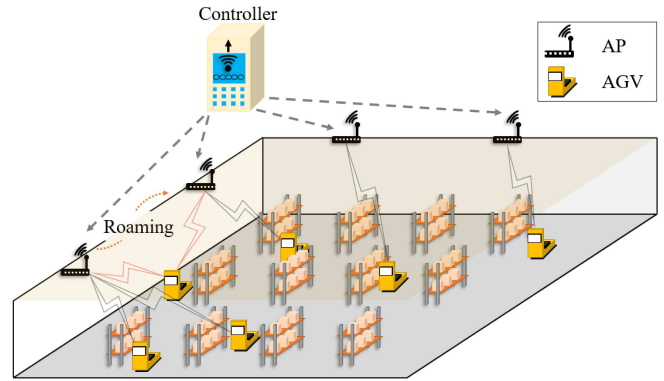}
    \caption{System architecture for 802.11ax based fast roaming in a smart warehouse.}
    \label{fig:env}
    \end{figure}

In recent years, learning methods have been proposed to alleviate the high complexity optimization required in conventional wireless communication methods \cite{14,15,16,m1,m2}. Reinforcement learning (RL) is one such model that optimizes learning weights based on environmental outcomes\cite{17}. However, traditional RL may not be suitable for high-dimensional and high-speed systems\cite{18}. Therefore, we propose using DQN approaches, which combine deep learning (DL) and RL, to expedite the extraction of valuable information from channel state information (CSI) and facilitate learning the best strategy. Unlike centralized methods and existing 802.11 standards, this approach aims to provide high efficiency and low latency, which have not been jointly considered in the existing literature \cite{5,6,10,11}. The main contributions of this paper are summarized as follows.
	\begin{itemize}
	
    \item We utilize the ns-3 network simulator with varying numbers of STAs and operating channels to emulate real-time Wi-Fi channel scanning and roaming based on the 802.11ax standard. Additionally, we collect CSI, including received signal strength indicator (RSSI), signal-to-noise ratio (SNR), and throughput, by receiving ProbeResponse frames from APs during each channel scan procedure, aiming to minimize latency during association.

	\item In adherence to the IEEE 802.11ax protocol, we design a distributed multi-agent deep Q-Learning for fast roaming (MADAR) algorithm, which learns to choose the best target AP for roaming. The distributed agents consist of collaborative macro-agents for STAs and competitive micro-agents for individual resources such as channels, resource units (RUs), and power. Our proposed MADAR scheme progressively learns the optimal strategy for AP assignment based solely on latency feedback.

	\item We evaluate the performance of the proposed MADAR scheme by simulating various numbers of serving STAs and operating APs in a Wi-Fi 6 scenario. By referencing practical datasets, we enhance the authenticity of the simulations. The MADAR scheme, in comparison to those in the open literature, demonstrates compatibility with operating channels across different transmission frequency bands and benefits from resilient allocation. Moreover, our proposed algorithm predicts a target AP that achieves low latency during STA association.

    \end{itemize}

    The remainder of this paper is organized as follows. Section II details the system model deployed by IEEE 802.11ax and formulates the problem of a roaming system. Section III elaborates on the proposed MADAR scheme, including epsilon-greedy and deep Q-learning algorithms, while Section IV presents performance evaluations. Finally, Section V draws the conclusions of this paper.

\section{System Model and Problem Formulation}

	\begin{figure}
       \centering
        \includegraphics[width=0.5 \textwidth]{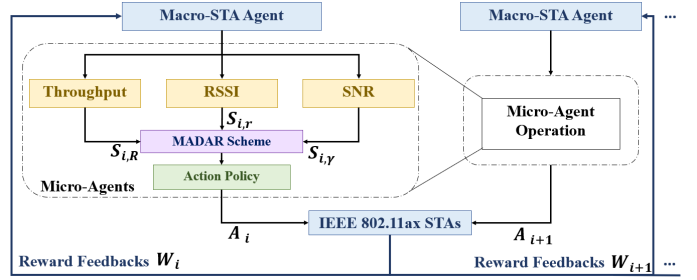}
        \caption{The proposed MADAR scheme. Distributed agents include collaborative multiple macro-agents of STAs and competitive micro-state-agents, i.e., throughput/RSSI/SNR.}
        \label{Fig.Multi-agent}
    \end{figure}

\subsection{System Model}   

As depicted in Fig. \ref{Fig.Multi-agent}, we consider a multi-STA network consisting of $M$ desired APs deployed with grid widths of $d_{m},\ \forall 1 \leq m\leq M$ in an environment using the 802.11ax standard. We assume that $N$ channels with bandwidth $B$ are available for these APs to scan. In the channel scanning phase, where the channel employs the OFDMA technique, the process of active scanning starts with the station initializing the $ProbeRequest$ frame, which includes the broadcast destination address, service set identifier (SSID), supported rate, and other details. The $K$ stations scan each channel, represented as $\mathcal{N} =\{1, 2,..., N\}$, sequentially from the current operating channel and send a $ProbeRequest$ frame to inquire about available APs on that channel. We then set two probe timers, $MinChannelTime$ and $MaxChannelTime$, to be $10$ and $100$ milliseconds, respectively. After the $k$-th station sends a $ProbeRequest$ management frame on the channel and waits for a $ProbeResponse$ frame, we observe the following:
\begin{itemize}
	\item If no AP responds within $MinChannelTime$, the station switches to the next channel and starts a new round of probing available APs.
	\item If the station receives any $ProbeResponse$ within $MinChannelTime$, the station waits to confirm whether there are other APs that have not responded until the end of $MaxChannelTime$.
\end{itemize}
Note that the probe timers $MinChannelTime$ and $MaxChannelTime$ are fixed in the system. The procedure is repeated until all channel scans are completed. By receiving the $ProbeResponse$ frame from the simulated network, we can obtain RSSI $r_{i, m}$ and SNR $\gamma_{i, m}$ of the $i$-th channel and $m$-th AP regarding the received frame signal quality from the physical layer. We then calculate the throughput of the $i$-th scanning channel after potential association and authentication, theoretically expressed by \cite{19} as
\begin{align}
\centering
R_{i, m} = \frac{E[P_{i, t}]}{E[L_{i, t}]}\leq R_{upp},
\end{align}
where $P_{i,t}$ denotes the payload information in a timeslot of the $i$-th channel, $L_{i,t}$ is the length of a slot time, and $R_{upp}$ is the achievable upper bound of Shannon capacity, which can be represented as
\begin{align}
\centering
R_{upp} = B\cdot\log_{2}\left(1+ \frac{\alpha_{i} p_{i} h_{i}}{\sigma^{2}_{i}}\right),
\end{align}
where $\alpha_{i}$ is an indicator for the selected channel $n$, $p_{i}$ is the transmission power for the selected channel $i$, $h_{i}$ is the channel gain derived from transmit power and collected RSSI, $\sigma^{2}_{i}$ is the noisy signal power including interference and noise power, and $B$ is the available operating bandwidth.

Moreover, we consider the overall handover delay formulated as $T = T_{CS}+T_{RO}+T_{cont}$, and $T_{CS}$ is the channel scanning time for $N$ channels, which can be represented as \cite{6}
\begin{equation}
\begin{aligned}
&T_{CS} = \sum_{i=1}^N T_{CS, i} = \sum_{i=1}^N {T_{pb, i}+T_{wt, i}+T_{sw, i}}
\\
&\quad\quad = \sum_{i=1}^N {T_{pb, i}+t_{min, i}+t_{max, i}+T_{sw, i}},
\end{aligned}
\end{equation}
where $T_{pb, i}$ denotes probe request time on channel $i$. $T_{wt, i}$ denotes the probe response waiting time, where $t_{min, i}$ is the minimum required response time of the $i$-th channel even with no existing AP, and $t_{max, i}$ is the maximum waiting time for all AP responses of the $i$-th channel on channel $i$. $T_{sw, i}$ denotes the switching time between channels. $T_{RO}$ is the roaming time, which can be formulated as
\begin{align}
\centering
T_{RO} = T_{au}+T_{as},
\end{align}
where $T_{au}$ denotes association delay and $T_{as}$ denotes authentication delay. In 802.11ax uplink, we need to implement UORA to contend for RUs. Then, $T_{cont}$ is the average contention delay for a station that successfully contends for the RUs, which can be expressed as
\begin{align}
T_{cont} =
\begin{cases}
T_{IFS}, & \mbox{if channel is idle,} \\
E[D_{s}] \times T_{IFS} + E[T_{BO}], & \mbox{otherwise,}
\end{cases}
\end{align}
where
\begin{align}
	&E[D_{s}] = \frac{1}{P_{s}} = \frac{1}{1-\left[1-\tau \left(1-p_{c}\right)\right]^{K_{sta}}},\\
	&E[T_{BO}] = \frac{(W_{0}-1)}{2} + \sum_{i=1}^U{(1-P_{s})^{i}\cdot \frac{(W_{0}\cdot 2^{i}-1)}{2}},
\end{align}
where $D_{s}$ means the number of stages elapsed until successful contention. $\tau$ is the STA transmit probability. $T_{IFS}$ is the interval frame space; when the channel is idle, it will wait for a while before detecting whether the channel is still idle. $T_{BO}$ means the sender will wait for a random backoff time after the channel is idle. $W_{0}$ means the minimum contention window size. $U$ is the maximum backoff stage. $P_{s}$ is the successful contention probability of at least one station successfully contending. $\tau$ is the transmission probability that an STA transmits a frame at the given UORA channel access contention. $p_{c}$ is the probability that at least one of the $(K_{sta}-1)$ remaining stations transmit in the selected RU, where $K_{sta}$ is the number of associated STAs contending to access RUs.

\subsection{Problem Formulation}
Our objective aims to minimize the total latency by considering channel scanning time $T_{CS}$, roaming time $T_{RO}$, and contention delay $T_{cont}$. The optimization problem can be formulated as
\begingroup
    \allowdisplaybreaks
    \begin{subequations}\label{OPT}
    \begin{align}
        \mathop{\min}_{\alpha_{i} b_{i} p_{i}} \quad &T = T_{CS}+T_{RO}+T_{cont}   \label{Obj}  
        \\
        \mathrm{ subject \, to }\quad &{\max R_{i}} \geq R_{thr}, \label{C1} &&\forall i, 
        \\
        & {\max r_{i}} \geq r_{thr}, &&\forall i, \label{C2} 
        \\
        & PER({\max s_{i}}) \leq \delta, &&\forall i, \label{C3} 
        \\
        & 0 \leq p_{i}  \leq  p_{max}, &&\forall i, \label{C4} 
        \\
        & \alpha_{i} \in \{0,1\}, &&\forall i, \label{C5}
        \\
        & \sum_{i=1}^N \alpha_{i} = 1, &&\forall i. \label{C6}                                   
    \end{align}
    \end{subequations}
    \endgroup
In $\eqref{C1}$ and $\eqref{C2}$, we optimize the scanned channel AP rate requirement and scanned signal quality, respectively. To guarantee QoS, in $\eqref{C3}$, we constrain the per-STA packet error rate (PER). $\eqref{C5}$ are generic binary set constraints, which represent AP association limitations and, when combined with $\eqref{C6}$, ensure that only a single channel can be selected to choose a target AP on the channel. In the following section, we intend to develop a deep Q Network model to minimize the total latency.

\section{Proposed Distributed Multi-Agent Deep Q-Learning for Fast Roaming Scheme (MADAR)} 

	\begin{figure}
       \centering 
        \includegraphics[width=0.45 \textwidth]{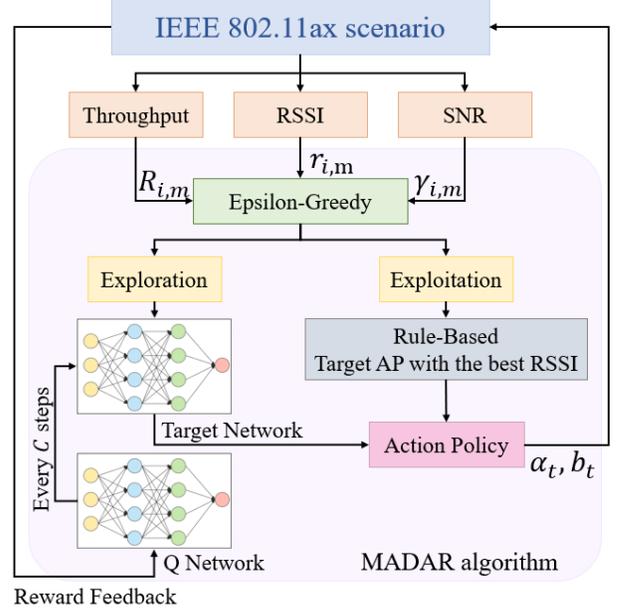}
        \caption{The architecture of Distributed Multi-Agent Deep Q-Learning and action policy training in the proposed MADAR scheme.}\label{Fig.algorithm}
    \end{figure}
    
According to existing literature, machine learning techniques are often used to predict channel state information, enabling flexible and dynamic associations with potential APs. RL is a type of machine learning in which an agent learns a task through a trial-and-error method by interacting with a dynamic environment. The proposed MADAR scheme employs a multi-agent DQN architecture, which is designed based on \emph{macro}-agents of STAs and \emph{micro}-agents of states. Notably, unlike the general state-driven RL, we use epsilon-greedy to select actions, taking advantage of exploitation and exploration during the training phase.

As illustrated in Fig. \ref{Fig.algorithm}, we collect the state $\mathcal{S} = \{S_{m, R}, S_{m, r}, S_{m, \gamma}\}$ from a real-time 802.11ax scenario. Note that $\mathcal{S}$ is a matrix representing the information of each candidate AP. In the action selection phase, we deploy epsilon-greedy using the parameter $\epsilon$. An epsilon-greedy agent may select an action $\mathcal{A} = \{\alpha_{t}, b_{t}\}$ with a sub-optimal state for low latency. Since our goal is to minimize delay, we allow this action to be learned by our network. Note that $\alpha_{t}$ is the number of nodes in the candidate AP, and $b_{t}$ is the operating channel of the target AP. As time passes and epsilon changes, we expect the action policy to gradually transition to the DQN after a converged training. In the learning phase, we consider four deep neural network (DNN) layers as our hidden layers with $\{16,32,64,128\}$ neurons, and we select $\tanh$ as our activation function in DNN layers. We fit the current state $\mathcal{S}_t$, next state $\mathcal{S}'_{t}$, action $a_t$, and reward ${W}_t$ into the DQN. Next, the gradient approach for the DQN optimizes the weight $\theta^{\mu}$, which is calculated as
\begin{align} \label{Q1}
	\nabla_{\theta^{ \mu}}{J} = E_{\{s, s', a, W\}}\left[ \nabla_{a} Q\left( s',a | \theta^{q}, s = s_{t},a = \mu\left(s|\theta^{\mu}\right)\right) \right].
\end{align}
The mean-square error (MSE) is selected as a loss function for the DQN expressed as
\begin{align} \label{Q2}
	L(\theta) = E_{\{s, a\sim\rho(s)\}} \left[y-Q(s,a|\theta)\right] ^{2},
\end{align}	
where $\rho(s)$ is a probability distribution over sequence $s$, and
\begin{align}
y = r+\gamma \max_{a'} \ {Q\left(s', a'\right)}
\end{align}
is the target reward in DQN. To determine which action yields the maximum reward, an agent must define the value of taking each action. The concrete procedure of MADAR scheme is demonstrated in Algorithm \ref{alg1}.

    \begin{algorithm}[!tb]
    \small
        \caption{\small Proposed MADAR Scheme}
        \SetAlgoLined
        \DontPrintSemicolon
        \label{alg1}
        \begin{algorithmic}[1]
        \STATE \textbf{Initialize:} Iteration index $t = 0, \epsilon, \theta^{\mu}$
        \STATE Set empty state table $\mathcal{S}$
        \WHILE{STA is operating}
        	\STATE $t \leftarrow t+1$
            \STATE Each STA attains state table when active scanning finishes 
                \STATE \textbf{(Exploration–Exploitation)}
                \IF {$random() < \epsilon$} 
                    \STATE Select $a_{i,t}, b_{i,t}$ having the optimal AP's RSSI from $S_{m,t}$
                \ELSE 
                    \STATE Select optimal prediction form target network by $\arg\max \,Q(s_{m,t},a_{m,t})$ 
                \ENDIF
                \STATE Interact with environments and obtain new state $S_{m,t+1}$ and reward $W_t$
                \STATE Memorize tuple of $\{S_{m,t}, S_{m,t+1}, a_{i,t}, W_t$ to replay memory buffer
                \STATE Train the Q network by $\eqref{Q1}$ and $\eqref{Q2}$
                \IF{$\mod (t,C)=0$}
                    \STATE Synchronize models of target network from Q network
                \ENDIF               
        \ENDWHILE    
      \end{algorithmic}
    \end{algorithm}

\section{Performance Evaluations}
    We evaluate the proposed MADAR algorithm by using ns-3 simulations to verify our proposed system model, we deploy five channels in different frequencies with three APs. With standard IEEE 802.11ax, we consider a Wi-Fi 6 network as specified in Table \ref{Parameter}, with three APs in different operating channels and $K = 20$ STAs, such that each STA is scattered randomly in area.  The STA needs to background active scan intervals is $200$ ms. Note that the channel scan processdure and the handoff threshold is $-85$ dB based on \cite{20}. The handoff process will be mandatory implement when the current link quality of the associated AP is less than handoff threshold. Because we need STAs to be able to move under different AP transmission ranges, we set the maximum transmission range of the AP to $30$ meters, and the distance between each AP is set to $25$ meters.
      
    \begin{table}[t]
    \centering
    \small
      \caption{Parameters of 802.11ax System}
     \begin{tabular}{l*{1}{l}r}
      \hline
      $\textbf{System Parameter}$ & $\textbf{Value}$ \\
      \hline
      Frequency & $2.4$ GHz, $5$ GHz\\
      Number of channels ($N$) & $5$ \\
      Number of AP ($M$) & $3$ \\
      Channel bandwidth ($B$) & $20$ MHz    \\
      Distance between APs  & $25$ meters \\
      Propagation max range & $30$ meters \\
      Interval of background active scan & $200$ ms\\
      MinChannelTime ($t_{min, i}$) & $10$ ms\\
      MaxChannelTime ($t_{max, i}$) & $100$ ms\\
      Association timeout & $100$ ms\\
      Learning rate ($\eta_1$) & $0.3$ \\  
      Exploration rate ($\epsilon$) & $0.3$ \\
      Handoff threshold & $-85$ dB \\
      Packet error rate ($\delta$) & $0.001$\\
      \hline
     \end{tabular} \label{Parameter}
    \end{table}
    
\begin{figure}
    	\centering
    	\subfigure[]{
    		\begin{minipage}{0.5\textwidth} 
                \includegraphics[width=\textwidth]{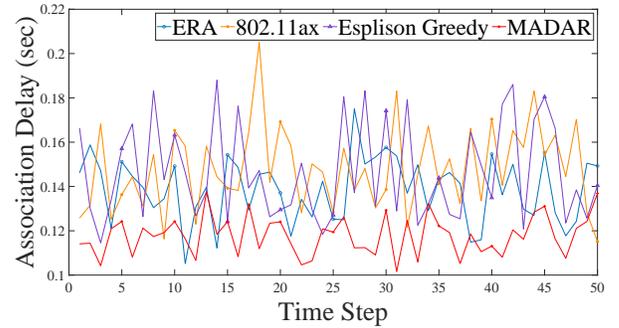} \\
    		\end{minipage}\label{Fig.assoc_delay_2_4}
    	}
    	\subfigure[]{
    		\begin{minipage}{0.5\textwidth}
    			\includegraphics[width=\textwidth]{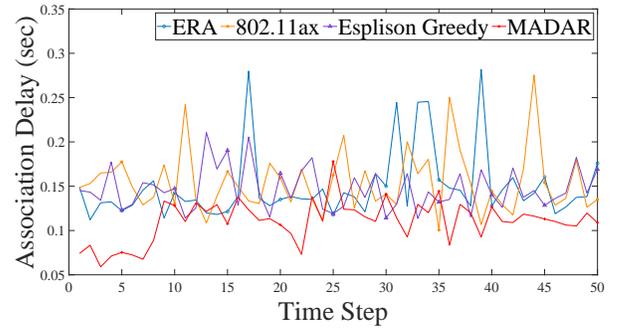} \\
    		\end{minipage}\label{Fig.assoc_delay_5}
    	}
    	\caption{Association delay performance of proposed MADAR compared with the benchmarks at channel bands of (a) 2.4 GHz and (b) 5 GHz.} 
    \end{figure}
    
In Figs. \ref{Fig.assoc_delay_2_4} and \ref{Fig.assoc_delay_5}, we evaluate the association delays $T_{RO}$ of the experiment using realistic AP setting (ERA), ns-3 network simulator (default 802.11ax protocol), epsilon-greedy, and our proposed MADAR scheme in different frequencies with $K=10$ STAs. Both the 802.11ax simulation and the ERA serve as our baseline, closely approximating real-world conditions. Once the neural network converges in our proposed scheme, we observe a higher tendency to select actions predicted by DQN rather than random selection. Note that the contention delay is included in each packet transmission time. As shown in Table \ref{Data of Assoc}, our simulated Wi-Fi system in ns-3 closely resembles the delay obtained from ERA. We compare this scenario with other benchmark methods. From the data in the table, we can see that although Epsilon-Greedy effectively reduces the association delay, its random selection during exploration may still choose the worst-performing AP. On average, if extreme values exist, the delay improvement may not be significant. In contrast, our proposed MADAR scheme uses the neural network model for prediction during target AP exploration, avoiding extreme values and yielding better delay reduction than conventional 802.11 and other methods.

    \begin{table}[!t]  
    \footnotesize
    \caption{Association Delay}
    \begin{subtable}[2.4 GHz]{ 
    \begin{tabular}{|c||c|c|c|c|c|}
        \hline
         Benchmark & ERA & 802.11ax & Epsilon-Greedy & MADAR \\ \cline{1-5}
	    Max &  0.175&	0.205&	0.188&	0.1375 \\ \hline
	    Min &  0.1052&	0.1151&	0.1145&	0.1016 \\ \hline
	    Avg &  0.1387&  0.148&  0.1466& 0.1178\\ \hline
    \end{tabular}
    }
    \end{subtable}
    \begin{subtable}[5 GHz] {
    \begin{tabular}{|c||c|c|c|c|c|}
        \hline
         Benchmark & ERA & 802.11ax & Epsilon-Greedy & MADAR \\ \cline{1-5}
	    Max &  0.2818&	0.2754&	0.2106&	0.178 \\ \hline
	    Min &  0.112&	0 .1005&	0.1135&	0.059 \\ \hline
	    Avg &  0.1492&  0.1545& 0.147& 0.1105\\ \hline
    \end{tabular}
    }
    \end{subtable}
    \label{Data of Assoc}
    \end{table}

    \begin{figure}
       \centering 
        \includegraphics[width=0.5\textwidth]{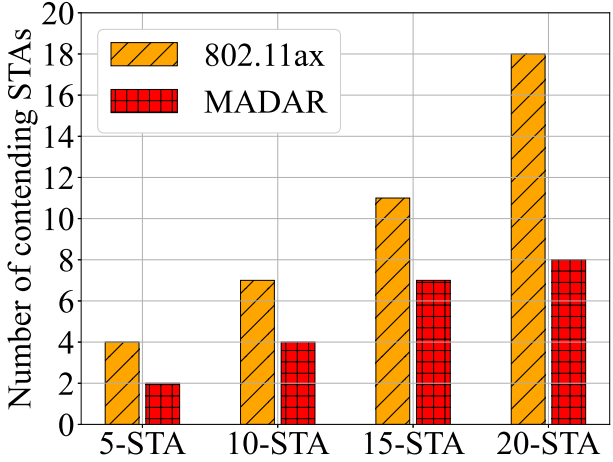} 
        \caption{The maximum number of supported contending STAs under different total numbers of operating STAs.}
        \label{Fig.STA_single_channel} 
    \end{figure}
    
Fig. \ref{Fig.STA_single_channel} shows the maximum contention STAs comparison for the proposed MADAR scheme and conventional 802.11ax under different numbers of STAs with $K = \{5, 10, 15, 20\}$. Conventionally, the target AP selection is based solely on RSSI, which potentially leads to channel congestion and increased packet collision probability if most STAs choose the AP with the best RSSI. Our proposed MADAR method dynamically learns from the delayed reward and CSI returned by the Wi-Fi system, effectively achieving network load balancing and reducing channel contention-induced delays.

    \begin{figure}
       \centering 
        \includegraphics[width=0.5\textwidth]{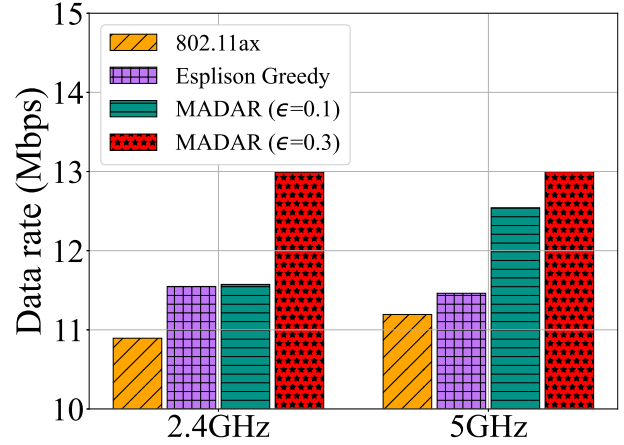}
        \caption{Data rate performance comparison for proposed MADAR scheme compared with benchmarks under different channel band 2.4 GHz and 5 GHz.}
        \label{Fig.data_rate}
    \end{figure}
    
Fig. \ref{Fig.data_rate} displays the average data rate for the proposed MADAR scheme, benchmarked against the Epsilon-Greedy method \cite{21} and conventional 802.11ax scheme. The Epsilon-Greedy method often chooses random APs, resulting in variable data rates in environments with a large number of STAs. Conventional 802.11ax has the worst performance in both frequency bands. Performance of MADAR varies with different epsilon values, as the epsilon parameter determines the proportion of DQN exploration (neural network) and exploitation (RSSI-based) at its early stage, eventually transitioning to full-DQN prediction. If epsilon is too small, useful actions may not be learned, leading to degraded results; if epsilon is too large, the learned features become random, which is similar to RSSI-based actions.
 
    \begin{figure}
       \centering 
        \includegraphics[width=0.5\textwidth]{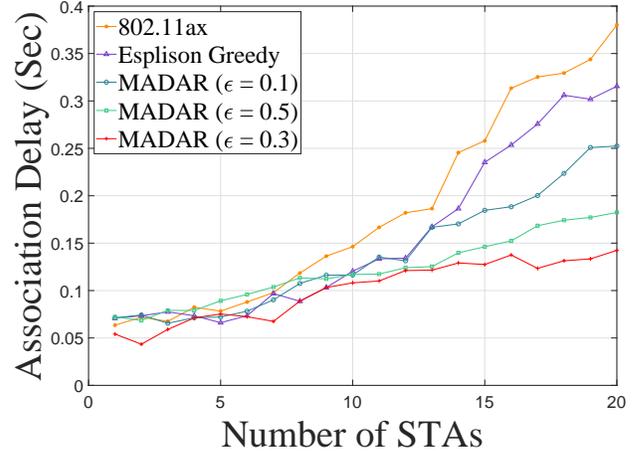} 
        \caption{Association Delay performance comparison for proposed MADAR scheme with conventional method under different operating numbers of STAs at 5 GHz.} 
        \label{Fig.num_vs_sta_assoc}
    \end{figure}

Fig. \ref{Fig.num_vs_sta_assoc} demonstrates the performance comparison for the proposed MADAR scheme with varying epsilon and the conventional roaming method under different numbers of STAs. In cases with a large number of STAs, channel congestion and a high probability of packet collision arise due to too many STAs in the same channel. As a result, association request must be retransmitted multiple times for successful roaming, which results in increasing association delay. The advantages of the proposed MADAR scheme can be observed, as it strikes a balance in loading between operating channels.

\section{Conclusions}
We have proposed the MADAR algorithm, which employs the deep Q-learning approach in an IEEE 802.11ax scenario to enhance Wi-Fi 6 roaming latency and rate through a decentralized control method. The MADAR-agent is designed to integrate the DQN and epsilon-greedy strategies, striking a compelling balance between exploration and exploitation by choosing between up-to-date and historical policies. Simulation results have demonstrated the potential benefits of the proposed MADAR algorithm in terms of network load balancing, packet error rate, and collision probability. By reducing dependence on APs or controllers, the MADAR algorithm, from the perspective of STA, is compatible with more devices supporting different standards and can enhance the potential benefits in the next-generation IEEE 802.11be standard, i.e., Wi-Fi 7.

\bibliographystyle{IEEEtran}
\bibliography{IEEEabrv}

\begin{thebibliography}{10}
\providecommand{\url}[1]{#1}
\csname url@samestyle\endcsname
\providecommand{\newblock}{\relax}
\providecommand{\bibinfo}[2]{#2}
\providecommand{\BIBentrySTDinterwordspacing}{\spaceskip=0pt\relax}
\providecommand{\BIBentryALTinterwordstretchfactor}{4}
\providecommand{\BIBentryALTinterwordspacing}{\spaceskip=\fontdimen2\font plus
\BIBentryALTinterwordstretchfactor\fontdimen3\font minus
  \fontdimen4\font\relax}
\providecommand{\BIBforeignlanguage}[2]{{%
\expandafter\ifx\csname l@#1\endcsname\relax
\typeout{** WARNING: IEEEtran.bst: No hyphenation pattern has been}%
\typeout{** loaded for the language `#1'. Using the pattern for}%
\typeout{** the default language instead.}%
\else
\language=\csname l@#1\endcsname
\fi
#2}}
\providecommand{\BIBdecl}{\relax}
\BIBdecl

\bibitem{1}
{Wi-Fi Alilance}, ``The {E}conomic {V}alue of {W}i-{F}i: A {G}lobal {V}iew
  (2021--2025),'' Sept. 2021.

\bibitem{2}
Z.~Zhang, Y.~Xiao, Z.~Ma, M.~Xiao, Z.~Ding, X.~Lei, G.~K. Karagiannidis, and
  P.~Fan, ``6{G} {W}ireless {N}etworks: {V}ision, {R}equirements,
  {A}rchitecture, and {K}ey {T}echnologies,'' \emph{IEEE Vehicular Technology
  Magazine}, vol.~14, no.~3, pp. 28--41, 2019.

\bibitem{acm}
L.-H. Shen, K.-T. Feng, and L.~Hanzo, ``{Five Facets of 6G: Research Challenges
  and Opportunities},'' \emph{ACM Computing Surveys}, vol.~55, no.~11, pp.
  1--39, 2023.

\bibitem{3}
S.~Bhattarai, G.~Naik, and J.-M.~J. Park, ``{U}plink {R}esource {A}llocation in
  {I}{E}{E}{E} 802.11ax,'' in \emph{Proc. IEEE International Conference on
  Communications (ICC)}, 2019, pp. 1--6.

\bibitem{4}
K.-H. Lee, ``{U}sing {O}{F}{D}{M}{A} for {M}{U}-{M}{I}{M}{O} {U}ser {S}election
  in 802.11ax-{B}ased {W}i{F}i {N}etworks,'' \emph{IEEE Access}, vol.~7, pp.
  186\,041--186\,055, 2019.

\bibitem{5}
A.~Mishra, M.~Shin, and W.~Arbaugh, ``{A}n {E}mpirical {A}nalysis of the
  {I}{E}{E}{E} 802.11 {M}{A}{C} {L}ayer {H}andoff {P}rocess,'' \emph{ACM
  SIGCOMM Computer Communication Review}, vol.~33, no.~2, pp. 93--102, 2003.

\bibitem{6}
L.~Hao, B.~Ng, and Y.~Qu, ``Self-optimizing {S}canning {P}arameters for
  {S}eamless {H}andover in {I}{E}{E}{E} 802.11 c,'' in \emph{Proc. IEEE
  Conference on Local Computer Networks (LCN)}, 2018, pp. 335--342.

\bibitem{7}
D.-J. Deng, Y.-P. Lin, X.~Yang, J.~Zhu, Y.-B. Li, J.~Luo, and K.-C. Chen,
  ``{I}{E}{E}{E} 802.11ax: {H}ighly {E}fficient {W}{L}{A}{N}s for {I}ntelligent
  {I}nformation {I}nfrastructure,'' \emph{IEEE Communications Magazine},
  vol.~55, no.~12, pp. 52--59, 2017.

\bibitem{8}
K.-H. Lee, ``{U}sing {O}{F}{D}{M}{A} for {M}{U}-{M}{I}{M}{O} {U}ser {S}election
  in 802.11ax-{B}ased {W}i{F}i {N}etworks,'' \emph{IEEE Access}, vol.~7, pp.
  186\,041--186\,055, 2019.

\bibitem{9}
K.~Kosek-Szott and K.~Domino, ``{A}n {E}fficient {B}ackoff {P}rocedure for
  {I}{E}{E}{E} 802.11ax {U}plink {O}{F}{D}{M}{A}-{B}ased {R}andom {A}ccess,''
  \emph{IEEE Access}, vol.~10, pp. 8855--8863, 2022.

\bibitem{m3}
L.-H. Shen, K.-H. Liao, and K.-T. Feng, ``{Queue-Aware Uplink Arbitration-based
  Contention and Downlink Resource Allocation for Multi-APs for IEEE 802.11ax
  WLANs},'' in \emph{Proc. IEEE Wireless Communications and Networking
  Conference (WCNC)}, 2022, pp. 1755--1760.

\bibitem{10}
M.~I. Sanchez and A.~Boukerche, ``{O}n {I}{E}{E}{E} 802.11k/r/v amendments:
  {D}o they have a real impact?'' in \emph{IEEE Wireless Communications},
  vol.~23, 2016, pp. 48--55.

\bibitem{11}
A.~Fink, R.~S. Mogensen, I.~Rodriguez, T.~Kolding, A.~Karstensena, and
  G.~Pocovi, ``{Empirical Performance Evaluation of Enterprise Wi-Fi for IIoT
  Applications Requiring Mobility},'' in \emph{Proc. IEEE European Wireless
  Conference}, 2021, pp. 1--8.

\bibitem{12}
Z.~Lin, Z.~Ni, L.~Kuang, C.~Jiang, and Z.~Huang, ``{Dynamic Beam Pattern and
  Bandwidth Allocation Based on Multi-Agent Deep Reinforcement Learning for
  Beam Hopping Satellite Systems},'' \emph{IEEE Transactions on Vehicular
  Technology}, vol.~71, no.~4, pp. 3917--3930, 2022.

\bibitem{13}
F.~Wilhelmi, B.~Bellalta, C.~Cano, and A.~Jonsson, ``{I}mplications of
  {D}ecentralized {Q}-{L}earning {R}esource {A}llocation in {W}ireless
  {N}etworks,'' in \emph{arXiv:1705.10508 [cs.NI]}, 2017.

\bibitem{14}
P.~V. Klaine, M.~A. Imran, O.~Onireti, and R.~D. Souza, ``{A} {S}urvey of
  {M}achine {L}earning {T}echniques {A}pplied to {S}elf-{O}rganizing {C}ellular
  {N}etworks,'' \emph{IEEE Communications Surveys \& Tutorials}, vol.~19,
  no.~4, pp. 2392--2431, 2017.

\bibitem{15}
Q.~Mao, F.~Hu, and Q.~Hao, ``{D}eep {L}earning for {I}ntelligent {W}ireless
  {N}etworks: {A} {C}omprehensive {S}urvey,'' \emph{IEEE Communications Surveys
  \& Tutorials}, vol.~20, no.~4, pp. 2595--2621, 2018.

\bibitem{16}
N.~C. Luong, D.~T. Hoang, S.~Gong, D.~Niyato, P.~Wang, Y.-C. Liang, and D.~I.
  Kim, ``{A}pplications of {D}eep {R}einforcement {L}earning in
  {C}ommunications and {N}etworking: {A} {S}urvey,'' \emph{IEEE Communications
  Surveys \& Tutorials}, vol.~21, no.~4, pp. 3133--3174, 2019.

\bibitem{m1}
L.-H. Shen, T.-W. Chang, K.-T. Feng, and P.-T. Huang, ``{Design and
  Implementation for Deep Learning Based Adjustable Beamforming Training for
  Millimeter Wave Communication Systems},'' \emph{IEEE Transactions on
  Vehicular Technology}, vol.~70, no.~3, pp. 2413--2427, 2021.

\bibitem{m2}
T.-W. Chang, L.-H. Shen, and K.-T. Feng, ``{Learning-Based Beam Training
  Algorithms for IEEE802.11ad/ay Networks},'' in \emph{Proc. IEEE Vehicular
  Technology Conference (VTC-Spring)}, 2019, pp. 1--5.

\bibitem{17}
R.~Ahmad, M.~D. Soltani, M.~Safari, A.~Srivastava, and A.~Das,
  ``{R}einforcement {L}earning {B}ased {L}oad {B}alancing for {H}ybrid {L}i{F}i
  {W}i{F}i {N}etworks,'' \emph{IEEE Access}, vol.~8, pp. 132\,273--132\,284,
  2020.

\bibitem{18}
V.~Aruna, L.~Anjaneyulu, and C.~Bhar, ``{D}eep-{Q} {R}einforcement {L}earning
  based {R}esource {A}llocation in {W}ireless {C}ommunication {N}etworks,'' in
  \emph{Proc. IEEE International Symposium on Smart Electronic Systems (iSES)},
  2022, pp. 66--72.

\bibitem{19}
H.~Wu, S.~Cheng, Y.~Peng, K.~Long, and J.~Ma, ``{IEEE 802.11 {D}istributed
  {C}oordination {F}unction (DCF): {A}nalysis and {E}nhancement},'' in
  \emph{Proc. IEEE International Conference on Communications (ICC)}, vol.~1,
  2002, pp. 605--609 vol.1.

\bibitem{20}
J.~Teng, C.~Xu, W.~Jia, and D.~Xuan, ``{D}-{S}can: {E}nabling {F}ast and
  {S}mooth {H}andoffs in {A}{P}-{D}ense 802.11 {W}ireless {N}etworks,'' in
  \emph{Proc. IEEE International Conference on Computer Communications
  (INFOCOM)}, 2009, pp. 2616--2620.

\bibitem{21}
A.~Idris, A.~Samaon, and M.~S. Idris, ``{P}erformance {A}nalysis of {R}esource
  {A}llocation {D}ownlink for {M}{I}{M}{O}-{O}{F}{D}{M}{A} {S}ystem using
  {G}reedy {A}lgorithm,'' in \emph{Proc. IEEE International Conference on
  Control System, Computing and Engineering (ICCSCE)}, 2016, pp. 157--162.

\end{thebibliography}
\end{document}